\documentclass[aps,pra,twocolumn,showpacs]{revtex4-1}
\usepackage{graphicx}

\begin{document}
\title{Quantum effects in the interaction of off-resonant coherent light with a single atom}

\author{Akihiro Yamaguchi and Holger F. Hofmann}
\email{hofmann@hiroshima-u.ac.jp}
\affiliation{
Graduate School of Advanced Sciences of Matter, Hiroshima University,
Kagamiyama 1-3-1, Higashi Hiroshima 739-8530, Japan}
\affiliation{JST,Crest, Sanbancho 5, Chiyoda-ku, Tokyo 102-0075, Japan
}

\begin{abstract}
Well controlled nonlinear interactions between light field pulses and single atoms could be used to implement optical quantum information technologies based on qubits encoded in superpositions of coherent states of light. Here, we investigate the transformation of a coherent light field input at a single atom sufficiently far from resonance to limit the decoherence effects associated with random excitations of the atom. The conditions for suppressing multi-photon scattering to implement arbitrarily large shifts of the coherent light amplitude without decoherence are studied. It is shown that quantum controlled coherent shifts can be achieved by sufficiently long coherent pulses, indicating the possibility of generating superpositions of coherent states with large amplitude differences. The  dominant multi-photon scattering effect associated with four wave mixing is also identified, and the spectral and temporal characteristics of the entangled photon pairs generated by this process are discussed. 
\end{abstract}


\maketitle

\section{Introduction}

Quantum optics has long been a testing ground for quantum information related technologies, since it combines the reliable control of field coherence by conventional linear optical elements with the possibility of precise measurements by photon detection. Ideally, a well-controlled optical non-linearity would complete the quantum optical toolbox by adding unconditional quantum gate operations based on photon-photon interactions. Initially, it was hoped that the interaction of light with a single atom in a microcavity would provide the solution, and initial experimental results did confirm that the non-linear effects observed in such a system would be sufficiently strong to realize interactions between individual photons \cite{Tur95}. However, these initial results were based on continuous wave input fields and their application to individual photons in well-defined finite time pulses is not immediately obvious. As the variety and the quality of technical implementations improves, the question of whether quantum information technologies could be based on non-linear optics has attracted renewed interest \cite{Mun05,Auf07,Mat09,Pro12,Li12,Ven13}. In particular, the fact that light field propagation effects need to be taken into account has been increasingly recognized, and the limitations of device fidelity caused by non-linear scattering in the time-frequency degree of freedom has been analyzed in a number of theoretical works \cite{Koj03,Hof03a,Hof03b,Kos04,Sha06,Ott06,Sha07,Nis09,Gea10,Mun10,He11,He12,Car12,Gea13}. 

Unfortunately, the present methods of analysis have not resulted in any simplifying intuitive insights, strengthening the impression that the light-matter interaction in free space is just too random and complex for any meaningful control at the quantum level \cite{Sha06,Sha07,Gea10}. It may therefore be helpful to take a step back, and to examine the way that the light-matter interaction is described in quantum mechanical models. What exactly is the cause of the difficulties, and might there be a way to find a more simple description of the interaction within a useful limit? 

Since the idea of non-linear switching works well in the classical limit, it has been suggested that the problem might be solved by using coherent states of light with high photon numbers \cite{Mun05,Pro12,Spi06}. In this limit, the non-linearity of off-resonant interactions between light and matter could be sufficient to cause the desired quantum effects without any additional decoherence caused by non-linear scattering between different frequencies. However, a more detailed theoretical analysis of the interaction between a coherent field and a single photon in a non-linear medium shows that it is difficult to avoid non-linear scattering effects between the different spatiotemporal modes describing the propagation of the light field through the medium \cite{Sha07,He11}. Nevertheless the robustness of optical coherence in the linear limit suggests that the absorption and re-emission of multiple photons will not necessarily result in decoherence. It may therefore be important to examine the quantum physics of photon absorption and emission in more detail, in order to distinguish coherent and incoherent optical effects at a more fundamental level. For this purpose, we here consider the interaction of a coherent input field with a single atomic system. Although this kind of coherently excited atomic system has been studied extensively using semiclassical models of the light-matter interaction, a complete quantum mechanical description of the output light field is non-trivial, since it generally involves quantum interferences between a large number of sequential photon absorptions and re-emissions. As we show in the following, the problem can be simplified considerably by analyzing the interaction terms that couple the light field to an atomic system. In particular, it is possible to show that a large number of absorption and re-emission events can be summarized by a simple coherent shift of the light field, corresponding to the effects of coherent atomic dipole emission. Significantly, we can show that this fully coherent dipole emission can involve a large number of photons, despite the fact that it is associated with the dipole of a single atomic system. It may therefore be possible to implement quantum controlled shifts of coherent states that can entangle two ground states of an atomic three level system with the two-dimensional Hilbert space formed by two nearly orthogonal coherent states of the light field. We also show that the effects of quantum noise associated with spontaneous absorption and emission events can be described by pairs of quantum jump events that generate pairs of photon states that are displaced in the optical phase space by the field amplitude of the coherent emission. These photon pairs can be described by an entangled two photon wavefunction similar to the two photon wavefunctions of down-converted photon pairs. It is pointed out that these two photon states describe the effects of squeezing associated with four wave mixing at the single atom non-linearity. Thus, the incoherent effects in the off-resonant light-matter interaction can be traced back to the inadvertent squeezing of the vacuum at frequencies far away from the original input frequency.

The rest of the paper is organized as follows. In section \ref{sec:model}, we introduce a simple model of one dimensional light field propagation and local interaction with an atomic system. We describe the propagation of light by time dependent operators and derive the effective Hamiltonian that describes the dynamics of the atom and its interaction with the field. In section \ref{sec:coherent}, the complex amplitude of the input coherent field is subtracted from the field operators to obtain effective field operators for whom the coherent input state is an effective vacuum. It is then possible to include the coherent input field in the atom dynamics, separating its effects from the effects of the light-matter interaction. In section \ref{sec:dress}, the interaction Hamiltonian is expressed in terms of eigenstates of the coherently driven atom and transitions between those eigenstates. All transitions between the eigenstates are then correlated with the emission of photons into the effective vacuum of the coherent input field. In section \ref{sec:shift}, it is shown that, in the absence of quantum jumps, the ground state of the coherently driven atom displaces the amplitude of the coherent field by a constant amount corresponding to the dipole of the ground state induced by the driving field. The conditions under which quantum jumps are sufficiently unlikely to be neglected are considered. Section \ref{sec:cat} discusses a possible application of large coherent shifts to a hybrid system of atomic qubits and qubits formed by superpositions of nearly orthogonal coherent states. Section \ref{sec:pairs} discusses the quantum states of the photon pairs emitted as a result of the quantum jumps between the dressed states of the driven atom. Section \ref{sec:concl} summarizes the results and concludes the paper.

\section{Light field propagation and interaction with a local atomic system}
\label{sec:model}

We consider the interaction between a light field propagating in one dimension and a single atom interacting with the field at a specific point, $x=0$. Note that this is different from the situation in a waveguide, where the light field can propagate in two directions, and the interaction with the atom may change the direction of propagation. In general, the one-dimensional description of light field propagation can be justified by identifying the absolute value $|x|=r$ with the distance from the atom, and the sign of $x$ with the propagation direction, so that incoming waves are described by $x<0$ and outgoing waves are described by $x>0$. As explained in \cite{Hof95}, this description can even be applied to the spherical waves of free space emission. For practical purposes, however, it will be helpful to consider a realization where the incoming and the outgoing fields propagate in a well focused beam shape along the axis of an optical cavity. The most simple realization of a well-controlled one-dimensional atom system would be a single atom in a one-sided cavity system, where almost 100\% of the spontaneous emission from the atom is emitted into the direction of the lower reflectivity mirror \cite{Hof03a}. Since the excitation of the atom is the time-reversed process of the emission, the incoming light must come from the same side of the cavity, so that $x<0$ describes incoming light at a distance of $r=-x$ from the cavity, while $x>0$ describes outgoing light at a distance of $r=x$. Significantly, the spatial degree of freedom $x$ does not simply describe the physical position, but combines information about the distance from the system with the direction of propagation. In this manner, incoming light can be separated from outgoing light, and all propagation effects can be described by a unidirectional shift of position with a constant velocity of $c$ in the positive $x$-direction. 

In the following, we will describe the light field in space and time by using the continuous modes defined by the annihilation operators $\hat{b}(x)$ of a photon at position $x$. For normalization purposes, the commutation relations of these operators are given by the Dirac delta function,
\begin{equation}
\label{eq:deltacom}
\left[ \hat{b}(x),\hat{b}^\dagger(x') \right] =\delta(x-x'). 
\end{equation}
The relevant transition of the atom can be described by the Pauli operators $\hat{\sigma}_i$ for the two level system defined by the ground state $\mid g \rangle$ and the excited state $\mid e \rangle$. Specifically, the operator $\hat{\sigma}_z$ describes the energy of the atom, with eigenvalues of $+1/2$ for the excited state and $-1/2$ for the ground state. The complex dipole of the atom is then described by the atomic annihilation operator,
\begin{equation}
\hat{\sigma}_- = \mid g \rangle\langle e \mid.
\end{equation}
An extension to multi-level systems is straightforward, requiring merely a proper labeling of the possible transitions to distinguish their effects on the light field. For the sake of simplicity, we will assume that there is only one relevant transition involved in the interaction, so that a two level system is adequate for the description of the atom. 

The total Hamiltonian of the light-matter interaction can be separated into three parts,
\begin{equation}
\hat{H}=\hat{H}_{\mathrm{light}} + \hat{H}_{\mathrm{int.}} + \hat{H}_{\mathrm{atom}}, 
\end{equation}
where $\hat{H}_{\mathrm{light}}$ describes the propagation of light in free space, $\hat{H}_{\mathrm{int.}}$ describes the interaction between the field and the atom, and  $\hat{H}_{\mathrm{atom}}$ describes the dynamics of the atomic system. We first simplify the problem by solving the propagation of light in free space in the Heisenberg picture, so that the propagation dynamics can be expressed by the time dependence of the annihilation operators,
\begin{equation}
\hat{b}(x,t)=\hat{b}(x - c (t-t^\prime), t^\prime).
\end{equation}
Effectively, light field propagation turns the spatiotemporal modes into elements of a quantum shift register, where the index $x$ at $t=0$ defines the time $t=-x/c$ at which the mode interacts with the atom. To simplify the notation, the operators of the light field can be labeled according to their arrival time at the atom,
\begin{equation}
\hat{b}(t)=\hat{b}(- c (t-t^\prime), t^\prime).
\end{equation}
In addition, we assume that the light field is described by a carrier frequency $\omega_{\mathrm{light}}$ that is detuned from the atomic resonance $\omega_{\mathrm{atom}}$ by 
\begin{equation}
\omega_{\delta}=\omega_{\mathrm{atom}}-\omega_{\mathrm{light}}.
\end{equation}
The coupling between the atom and the light field can be described by a local interaction with the mode $\hat{b}(t)$, where the coupling coefficient iBrother トナーカートリッジ　Magentas given by the dipole relaxation rate $\Gamma$, which is equal to one half of the spontaneous emission rate of the atom \cite{Hof03a,Hof95}. The effective time-dependent Hamiltonian describing the light-matter interaction can then be written as
\begin{equation}
\label{eq:Heff}
\frac{1}{\hbar} \hat{H}_{\mathrm{eff}} = \omega_\delta \hat{\sigma}_z -\sqrt{2\Gamma c} \left(\hat{b}(t)\hat{\sigma}_-^\dag+\hat{b}^\dag(t)\hat{\sigma}_- \right). 
\end{equation}
This Hamiltonian describes the time evolution of the quantum state, where $\hat{b}(t)$ indicates that the effects of the interaction act on different parts of the multi-mode light field as the light field propagates past the atom. In principle, this operator can be used to formulate the Schroedinger equation in the photon number basis, and the output state of the interaction dynamics could be found by integrating the quantum coherent superpositions of all possible sequences of absorption and re-emission, as was done for the two photon case in \cite{Koj03}. However, this approach is somewhat impractical in the case of a strong coherent input field due to the large number of possible absorptions and re-emissions associated with the high photon number in the input. In the following, we will therefore adapt the formulation of the interaction to the specific situation of a coherent input field in order to identify the essential physics of the interaction in the context of a strong off-resonant driving field. 

\section{Coherent state input as an effective vacuum}
\label{sec:coherent}

Coherent states correspond to a vacuum state displaced from the zero field by an amplitude of $\alpha$, which corresponds closely to the field amplitude used in semiclassical theories of the light-matter interaction. In the multimode case, the coherent state is defined by the amplitudes in each mode, as given by the eigenvalues of the annihilation operators. For continuous time dependent fields
\begin{equation}
	\hat{b}(t) \mid \beta(t) \rangle = \frac{1}{\sqrt{c}}\beta(t) \mid \beta(t) \rangle,
\end{equation} 
where $\beta(t)$ is normalized so that $|\beta(t)|^2$ gives the average rate of photons incident on the atom at time $t$. Using all of the amplitudes $\beta(t)$, it is possible to define a multi-mode displacement operator $\hat{D}_\beta$, such that the coherent state can be written as a displaced vacuum,
\begin{equation}
\mid \beta(t) \rangle = \hat{D}_\beta \mid \mbox{vac.} \rangle.
\end{equation} 
Thus, the coherent state can be treated as an effective vacuum with regard to a zero-point field value of $\beta(t)$. The difference between this zero-point field and the quantum mechanical field described by the annihilation operators $\hat{b}(t)$ can be represented by the effective field operators
\begin{eqnarray}
\hat{b}_\mathrm{eff}(t) &=& \hat{D}_\beta \hat{b}(t) \hat{D}_\beta^\dagger
\nonumber \\ &=&
\hat{b}(t)-\frac{1}{\sqrt{c}}\beta(t).
\end{eqnarray}  
The effective field operators $\hat{b}_\mathrm{eff}(t)$ describe the light-matter interaction in terms of photon absorption and emission events relative to the effective vacuum defined by $\mid \beta(t) \rangle$. In particular, the application of the effective creation operator $\hat{b}^\dagger_\mathrm{eff}(t)$ to the initial coherent state describes the generation of an effective single photon state,
\begin{eqnarray}
\label{eq:delta}
\mid \Delta(t^\prime) \rangle &=& \hat{b}_{\mathrm{eff}}^\dagger(t^\prime) \mid \beta(t) \rangle 
\nonumber \\ &=& \hat{D}_\beta \hat{b}^\dagger(t^\prime) \mid \mbox{vac.} \rangle.
\end{eqnarray}
As the second line of Eq. (\ref{eq:delta}) shows, this state can also be obtained by displacing the state of a single photon at time $t^\prime$ by the coherent amplitudes $\beta(t)$. Likewise, a series of multiple effective emissions at different times $t^\prime$ can be expressed by the corresponding coherently displaced multi-photon states. 

By describing the quantum state of the light field in terms of displaced photon number states, we can summarize the effects of the coherent field amplitude on the atom and express the remaining interaction in terms of a limited number of additional photon emissions caused by transitions between the excited state and the ground state of the atom. The Hamiltonian describing the dynamics can then be expressed as
\begin{eqnarray}
\label{eq:drivingH}
\frac{1}{\hbar} \hat{H}_{\mathrm{eff}} &=& \omega_\delta \hat{\sigma}_z -\sqrt{2\Gamma} \left(\beta(t) \hat{\sigma}_-^\dagger+\beta^*(t)\hat{\sigma}_- \right)
\nonumber \\ &&
-\sqrt{2\Gamma c} \left(\hat{b}_{\mathrm{eff}}(t) \hat{\sigma}_-^\dagger+\hat{b}_{\mathrm{eff}}^\dagger(t)\hat{\sigma}_- \right). 
\end{eqnarray} 
Significantly, the part of the light-matter interaction that describes the effect of the coherent driving field on the atom is now independent of the quantum state of the light field and can be treated as part of the internal dynamics of the atom. It is therefore possible to summarize a large part of the interaction dynamics by including the semiclassical response of the atom to the field in the description of the atom - a method also known as ``dressing'' the atom in the field. However, the fact that the atom is usually in a partially excited state means that there can still be a large number of effective emissions of displaced photon states into the effective vacuum. To keep track of these additional emissions, it is useful to consider the dynamics of the atom in the presence of (or ``dressed'' in) a coherent driving field.

\section{Dressed states of the atom}
\label{sec:dress}

If we assume that the coherent field changes only slowly, the semiclassical part of the Hamiltonian can be diagonalized to obtain dressed state solutions for stationary states of the coherently driven atom. Specifically, the atomic Hamiltonian is now proportional to a Pauli operator associated with an axis tilted by an angle of $\theta$ from the original $z$-direction,
\begin{equation}
\hat{\sigma}_z^\prime = \cos \theta \hat{\sigma}_z + \sin \theta \frac{1}{2}
\left(e^{-i \phi} \hat{\sigma}_- + e^{i \phi} \hat{\sigma}_-^\dagger \right),
\end{equation}
where $\phi$ is the phase of the field amplitude $\beta(t)$ and $\theta$ depends on the ratio of amplitude $|\beta(t)|$ and detuning $\omega_\delta$ according to
\begin{equation}
\label{eq:theta}
\theta = \arctan\left(\frac{2\sqrt{2\Gamma}}{\omega_\delta} |\beta(t)|\right).
\end{equation}
The change to the dressed state basis also modifies the description of transitions between the eigenstates of the atomic Hamiltonian. These transitions can be represented by a transformed atomic annihilation operator,
\begin{eqnarray}
\hat{\sigma'}_- &=& \cos^2 \left(\frac{\theta}{2}\right) \hat{\sigma}_-
- e^{i 2 \phi} \sin^2 \left(\frac{\theta}{2}\right) \hat{\sigma}_-^\dagger
\nonumber \\ && \hspace*{2cm}
- e^{i \phi} \sin \theta \hat{\sigma}_z.
\end{eqnarray}
With this transformation, it is possible to express the interaction Hamiltonian in terms of transitions between the dressed states of the atom. The result reads
\begin{eqnarray}
\label{eq:dressedH}
\lefteqn{\frac{1}{\hbar} \hat{H}_{\mathrm{eff}} =}
\nonumber \\ &&
 \omega_\beta \hat{\sigma}_z^\prime -\sqrt{2\Gamma c} \sin \theta \left(e^{-i \phi} \hat{b}_{\mathrm{eff}}(t)+e^{i \phi} \hat{b}_{\mathrm{eff}}^\dagger(t) \right) \hat{\sigma}_z^\prime
\nonumber \\ &&
-\sqrt{2\Gamma c} \cos^2 \left(\frac{\theta}{2}\right) \left(\hat{b}_{\mathrm{eff}}(t) \hat{\sigma'}_-^\dagger+\hat{b}_{\mathrm{eff}}^\dagger(t)\hat{\sigma'}_- \right)
\nonumber \\ &&
-\sqrt{2\Gamma c} \sin^2 \left(\frac{\theta}{2}\right) \left(\hat{b}_{\mathrm{eff}}^\dagger(t) \hat{\sigma'}_- ^\dagger+\hat{b}_{\mathrm{eff}}(t)\hat{\sigma'}_- \right).
\end{eqnarray} 
In this formulation of the Hamiltonian, it is possible to separate the dynamics into a time evolution that conserves $\hat{\sigma}_z^\prime$ and a series of quantum jumps between the eigenstates of $\hat{\sigma}_z^\prime$ that are correlated with discontinuous changes in the light field represented by the creation of a photon in the effective vacuum described by the coherent field. Since the quantum jumps result in the creation of photon states in the effective vacuum, the propagation of these photon states away from the atom leaves an irreversible record of the quantum jump in the emitted field. Although the process is coherent and the output state should be expressed by a superposition of different quantum jump times, the fact that the output state components can always be distinguished by the number of photons effectively created in the output field means that it is convenient to expand the solution in terms of this number, starting from the solution for zero quantum jumps. If this series converges, it is possible to integrate the Schroedinger equation by considering only the lowest relevant numbers of quantum jumps in the solution, similar to the Feynman path approach to interactions between elementary particles.

In a specific scenario, the input light field $\beta(t)$ will be time dependent. However, we can assume that the time dependence is sufficiently slow compared to the detuning dynamics described by $\omega_\delta$, so that the quantum state of the atomic system will follow the changes of the eigenstates of $\sigma_z^\prime(t)$ adiabatically. If the atom is initially in the ground state, it will remain in the instantaneous groundstate $\mid g^\prime \rangle$ unless a quantum jump takes it to the excited state. Without a driving field, such quantum jumps would be impossible, since the excitation of the atom would require the annihilation of a photon, and no such photon is available in the input state. However, the presence of the driving field makes such quantum jumps from ground state to the excited states possible, as expressed by the product of the creation operators $\hat{b}_{\mathrm{eff}}^\dagger(t) \hat{\sigma'}_-^\dagger$ in the Hamiltonian. This operator results in a transition to the excited state and a simultaneous creation of a photon in the effective vacuum represented by the coherent input field. The probability of such a quantum jump for an infinitesimal time interval $dt$ is given by the interaction coefficient in the Hamiltonian. For the transition from ground state to excited state, the probability per time interval is
\begin{equation}
\label{eq:jumprate}
\frac{d P_{\mathrm{jump}}(g^\prime \to e^\prime)}{dt} = 2 \Gamma \sin^4 \left(\frac{\theta}{2}\right).
\end{equation}
Note that $2 \Gamma$ is the spontaneous emission rate of the excited state in the absence of a driving field. For driving fields that are sufficiently weak compared to the detuning ($\theta \ll 1$), the rate of transitions is proportional to the fourth power of the field amplitude $|\beta(t)|$. In this limit, we can expect the atom to remain in the dressed ground state $\mid g^\prime \rangle$ during most of the dynamics, with only a small number of quantum jumps to the excited state. 

Once the atom is in the excited state due to a quantum jump, the probability of another quantum jump back to the ground state is given by  
\begin{equation}
\label{eq:relax}
\frac{d P_{\mathrm{jump}}(e^\prime \to g^\prime)}{dt} = 2 \Gamma \cos^4 \left(\frac{\theta}{2}\right).
\end{equation}
In the limit of weak driving fields ($\theta \ll 1$), this probability is equal to the spontaneous emission rate of the atom. The atom will therefore return to the dressed ground state within a time comparable to the lifetime of the excited state in the absence of a driving field. 

In general, the response of the atom to the coherent driving field is given by a coherent superposition of all possible combinations of quantum jumps. Significantly, each combination of quantum jumps is distinguishable in the output state of the light field, since the excitation of a photon in the effective vacuum represented by the input state transforms the state of the light field to an orthogonal state. Different quantum jump sequences are therefore represented by orthogonal states of the output light field, and no integration over paths with indistinguishable outcomes is necessary.

\section{Coherent shift of the output field}
\label{sec:shift}

The Hamiltonian given in Eq. (\ref{eq:dressedH}) acts on the light field in two different ways. In addition to the quantum jumps between the eigenstates of the dressed states, there is also the interaction represented by the product of $\hat{\sigma}_z^\prime$ and the field operators. If we assume that no quantum jumps occur and the atom is always in its ground state $\mid g^\prime \rangle$, the approximate Hamiltonian of the light-matter interaction is given by
\begin{equation}
	\frac{\hat{H}_0}{\hbar}=-\frac{\omega_\delta}{2} + \frac{1}{2}\sqrt{2 \Gamma c} \sin \theta \left(e^{-i \phi} \hat{b}_{\mathrm{eff}}(t)+e^{i \phi} \hat{b}_{\mathrm{eff}}^\dagger(t) \right).
\end{equation}
This Hamiltonian acts on the light field as it passes the atom. Specifically, it generates a displacement of the state that changes the amplitude of the coherent input state by an amount proportional to the coefficients in the Hamiltonian. Using Eq.(\ref{eq:theta}), this coefficient can be expressed in terms of the input field amplitude,
\begin{equation}
\frac{1}{2}\sqrt{2 \Gamma} \sin \theta e^{i \phi} = \frac{2 \Gamma}{\sqrt{\omega_\delta^2+8 \Gamma |\beta(t)|^2}} \; \beta(t). 
\end{equation}
The non-linear dependence of the coefficient on $\beta(t)$ reflects the saturation effects in the excitation of the atom by the driving field. In terms of the timescales of the atom-field dynamics, the condition for weak excitation ($\theta \ll 1$) is given by $\omega_\delta \gg 8 \Gamma |\beta(t)|^2$, in which case the approximate value of the coherent shift induced by the atom in the optical field is given by the semiclassical linear response,
\begin{equation}
\beta_{\mathrm{out}}(t) = \beta(t) - i \frac{2 \Gamma}{\omega_{\delta}} \beta(t).
\end{equation}
In the limit of weak excitation by the off-resonant driving field, the response of the single atom can therefore change the coherent amplitude of the input light in accordance with the semiclassical theory, without any decoherence of the light field. 

Significantly, the coherent shift induced by the single atom can involve a large number of photons. For a rectangular pulse with constant amplitude $\beta(t)=\beta_0$, the average photon number in the pulse is given by $|\beta_0|^2 T$, where $T$ is the duration of the pulse. Therefore, the coherent state amplitude of the total pulse is $\alpha_{\mathrm{in}}=\beta_0 \sqrt{T}$, and the coherent shift induced by the single atom is
\begin{equation}
\label{eq:shift}
\alpha_{\mathrm{out}}-\alpha_{\mathrm{in}} = 
- i \frac{2 \Gamma}{\omega_{\delta}} \beta_0 \sqrt{T}.
\end{equation}
By making the pulse time arbitrarily long, a single atom can thus induce an arbitrarily large coherent shift. This means that the quantum state of a single atom could be used to control the amplitude of coherent states, permitting the realization of atom-light quantum gates that could encode quantum information into superpositions of nearly orthogonal coherent state. However, it is necessary to ensure that the amplitude $\beta_0$ is indeed low enough to neglect the effects of quantum jumps during the pulse time $T$. According to Eqs.(\ref{eq:theta}) and (\ref{eq:jumprate}), the rate of quantum jumps for an amplitude of $\beta_0$ is
\begin{equation}
\label{eq:jump2}
\frac{d P_{\mathrm{jump}}(g^\prime \to e^\prime)}{dt} = \frac{1}{2 \Gamma} \left|2 \frac{2 \Gamma}{\omega_\delta} \beta_0 \right|^4.
\end{equation}
If the amplitude $\beta_0$ is expressed in terms of the total coherent shift given in Eq.(\ref{eq:shift}), the total probability of a quantum jump during the pulse time $T$ can be given as
\begin{equation}
\label{eq:Pjump}
P_{\mathrm{jump}} \approx \frac{|2 (\alpha_{\mathrm{out}}-\alpha_{\mathrm{in}})|^4}{2 \Gamma T}.
\end{equation}
To achieve coherent shifts larger than one, it is therefore necessary to make the duration of the input pulses sufficiently longer than the spontaneous emission lifetime of the excited state. 

Note that there is no theoretical limit to the amount of coherent shift that can be obtained at negligible quantum jumps. However, the coherence of the atom and the coherence of the input pulse must be stable over the pulse time $T$. For typical atomic transitions at optical freuqencies, $1/\Gamma$ will be several nanoseconds long, so it is desirable to achive coherence times in the microsecond regime in order to suppress quantum jump effects at sufficiently large coherent shifts. According to Eq.(\ref{eq:Pjump}), the coherent shift that can be obtained with $\Gamma T \approx 10^3$ and a quantum jump probability of $P_{\mathrm{jump}}=0.1$ is still only $\alpha_{\mathrm{out}}-\alpha_{\mathrm{in}}\approx 1.9$. In practice, it may be difficult to push the limit much farther due to the difficulty of achieveing long coherence times. This result thus illustrates the technical challenge involved in controlling macroscopic coherent shifts through the microscopic degrees of freedom of a single atom.

\section{Quantum control of coherent shifts by ground state superpositions}
\label{sec:cat}

\begin{figure}[th]
\begin{picture}(240,180)
\put(0,0){\makebox(240,180){\vspace{-2cm}\hspace{-0.8cm}
\scalebox{0.5}[0.5]{
\includegraphics{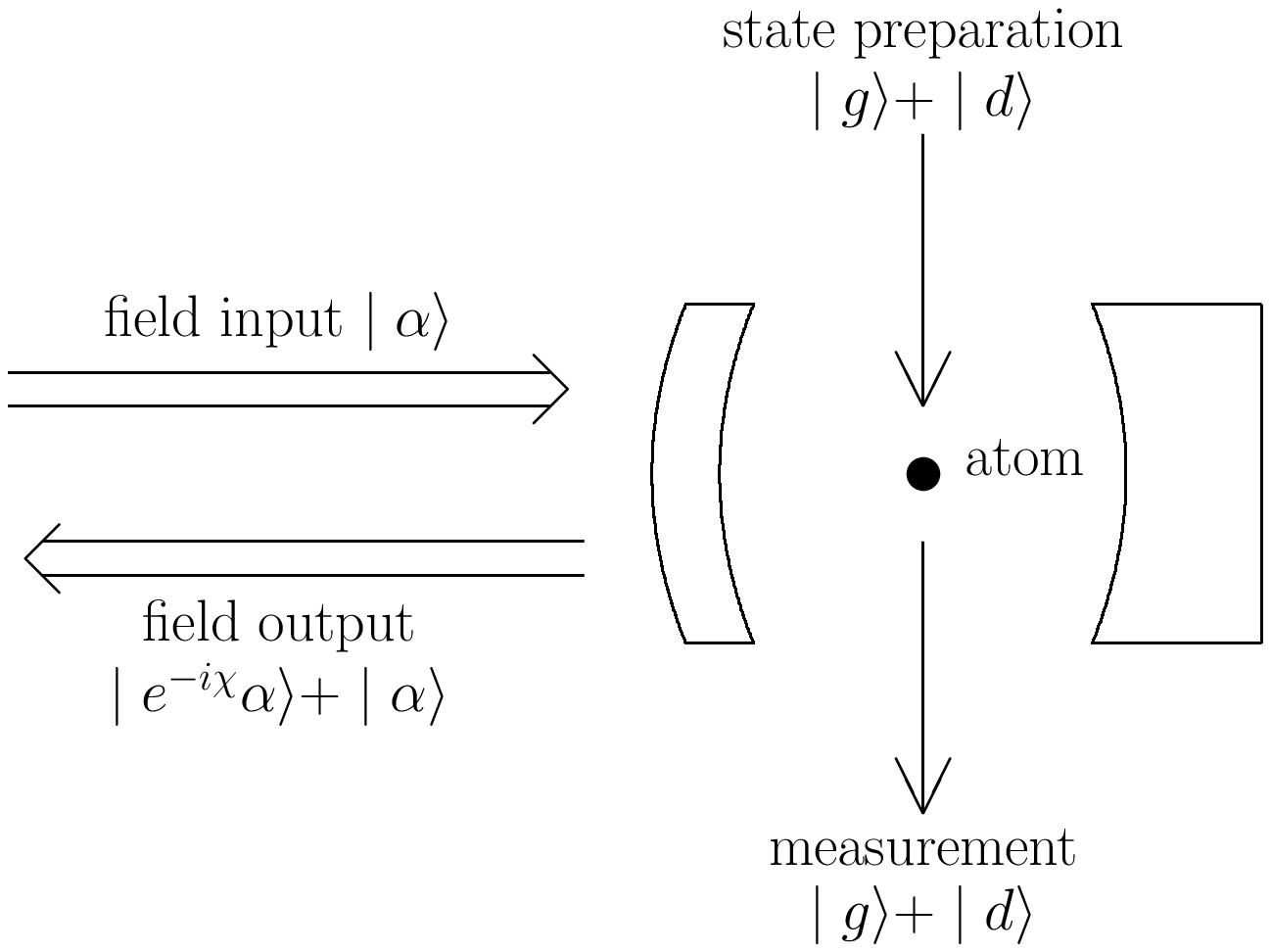}}}}
\end{picture}
\caption{\label{fig1} Illustration of an experimental realization of a superposition of coherent states with different amplitudes. A three level atom is prepared in an equal superposition of its ground state $\mid g \rangle$ and a state $\mid d \rangle$. It is then dropped through the one sided cavity, where the state $\mid g \rangle$ induces a phase shift $\chi$ in the incident light field $\mid \alpha \rangle$, while the state $\mid d \rangle$ has no effect. Finally, the atom exiting the cavity is detected in an equal superposition of $\mid g \rangle$ and $\mid d \rangle$, resulting in the desired superposition in the optical output.}
\end{figure}

One exciting possibility that emerges from the observation that a single atom can induce arbitrarily large coherent shifts in its linear interaction with an input field is the control of coherent state amplitudes by quantum states of the atom. In particular, it may be possible to implement a quantum controlled shift gate, if the interaction between the atom and the light field depends on the quantum state of the atom \cite{Spi06}. For simplicity, we will consider a three level atom described by the states $\mid g \rangle$, $\mid e \rangle$, and $\mid d \rangle$, where the third state does not interact with the light field at all. The atom can now be used as a qubit, where information can be encoded in $\mid g \rangle$ and $\mid d \rangle$. To prepare and measure the atom, it may be useful to drop it through the cavity, using strong pump pulses for preparation and measurement before and after it passes through the cavity to interact with the one-dimensional field incident on the face of the cavity. A schematic setup for such an experiment is shown in Fig. \ref{fig1}. While the atom is in the cavity, it interacts with the coherent input pulse according to the theory discussed above. If $P_{\mathrm{jump}} \ll 1$, the total effect of the interaction will be a conditional phase shift, 
\begin{eqnarray}
\hat{U} \mid \alpha \rangle \otimes \mid d \rangle &=&  \mid \alpha \rangle \otimes \mid d \rangle
\nonumber \\
\hat{U} \mid \alpha \rangle \otimes \mid g \rangle &=&  \mid e^{-i \chi} \alpha \rangle \otimes \mid g \rangle,
\end{eqnarray}
where $\chi=2 \Gamma/\omega_\delta \ll 1$. For sufficiently large coherent amplitudes $\alpha$, the coherent states in the output have negligible overlap. It is then possible to generate superpositions of distinguishable coherent states by preparing the atom in an equal superposition of $\mid g \rangle$ and $\mid d \rangle$, interacting the coherent light with the atom, and performing a projective measurement of an equal superposition of $\mid g \rangle$ and $\mid d \rangle$ after the interaction. The final field state after this operation reads
\begin{equation}
\left(\langle g \mid + \langle d \mid\right) \hat{U} \mid \alpha \rangle \otimes \left(\mid g \rangle + \mid d \rangle\right) =
\mid e^{-i \chi} \alpha \rangle + \mid \alpha \rangle.
\end{equation}
The interaction between strong coherent light and a single atom could therefore be used to realize a cat-state superposition of coherent states with negligible overlaps. Oppositely, evidence for such a non-classical coherence in the output field can be used to confirm whether a quantum controlled shift gate has been realized or not.

The present scenario suggests that a possible solution to the problems raised by uncontrolled non-linear scattering in the spatiotemporal degrees of freedom of the light field might be a hybrid architecture of atoms and fields, where the linear interaction between the light field and the atom is controlled by the quantum state of the atom. Importantly, the results derived above show that this approach may even work if the response of the atom induces very large coherent shifts. However, this approach assumes that the quantum jumps that are also generated by the Hamiltonian in Eq. (\ref{eq:dressedH}) can be neglected. In practice, this may require a careful trade-off between the coherent shift and the errors caused by the quantum jumps, as expressed by Eq.(\ref{eq:Pjump}). Note that quantum jumps only occur if the atom is in the $\mid g \rangle$ state, so that a single quantum jump immediately identifies the state and destroys the coherence between $\mid g \rangle$ and $\mid d \rangle$. The generation of a superposition of coherent states therefore requires that $P_{\mathrm{jump}}\ll 1$. However, the quantum jumps themselves may have interesting and potentially useful non-classical properties of their own, so it may be important to understand the properties of these non-linear contributions to the response of the atom in more detail.

\section{Photon pair emission by transitions between dressed states}
\label{sec:pairs}

As discussed in section \ref{sec:dress}, a quantum jump that takes the atom from its dressed ground state $\mid g^\prime \rangle$ to the dressed excited state $\mid e^\prime \rangle$ will be followed by a quantum jump back to the ground state within a time that remains close to the original lifetime of the excited state without driving field. In the limit of weak excitation ($\theta \ll 1$), these quantum jump pairs are approximately independent of each other, since the atom will have relaxed back to the ground state long before the next quantum jump initiates another pair event. 

In the light field, each quantum jump is represented by the creation of a photon in the effective vacuum. In addition, the sign of the displacement is reversed during the time interval between the quantum jumps. However, the effect of this displacement is directly related to the value of $\theta$ and can be neglected for $\theta \ll 1$. We can therefore describe the effects of a quantum jump event in the output of the light field in terms of the wavefunction of a photon pair generated in the effective vacuum represented by the coherent field. The total probability density for the photon pairs is given by Eq.(\ref{eq:jumprate}). The probability distribution over the time difference $|t_2-t_1|$ between the initial quantum jump and the final quantum jump is given by an exponential function with a relaxation rate of $2 \Gamma^\prime=2 \Gamma \cos^4(\theta/2)$, as given by Eq.(\ref{eq:relax}). During the time interval between the jumps, the phase dynamics of the excited state applies, so that the total phase of the two photon wavefunction is given by $-\omega_\beta |t_2-t_1|$. With this information, it is possible to write down the two photon wavefunction that describes the effects of a quantum jump event on the output state,
\begin{equation}
\label{eq:timepair}
\psi(t_1,t_2) = \sqrt{2} \Gamma^\prime \tan^2\left(\frac{\theta}{2}\right)
 \exp\left(-(\Gamma^\prime + i \omega_\beta) |t_2-t_1| \right). 
\end{equation}
Since the two photons are indistinguishable, the definition of $t_1$ and $t_2$ is rather arbitrary. However the absolute value of the difference ensures that the later time appears with a positive sign, and the earlier time appears with a negative sign in $|t_2-t_1|$. Interestingly, this means that there is a correlation between photon arrival time and photon frequency. The frequency of the first photon is reduced by the difference between atomic frequency and the input frequency, while the frequency of the second photon is increased to match the atomic frequency. This suggests a fairly simple picture of the non-linear scattering process in time, where two photons arrive simultaneously at the atom, and one photon is absorbed resonantly by taking the necessary energy difference from the other photon. Since the photon that has supplied this energy difference is not absorbed, it arrives at the detectors earlier, while the absorbed photon is emitted only after a time delay corresponding to the spontaneous emission time. 

It is also possible to interpret the pair emission as a squeezing effect associated with the quantum limit of four wave mixing at a single atom. Since the photons are emitted at frequencies that are quite different from the input frequency $\omega_{\mathrm{light}}$, the output state is a squeezed vacuum, where the quadratures of the frequencies near the atomic resonance $\omega_{\mathrm{atom}}$ are entangled with the quadratures of $2 \omega_{\mathrm{light}}-\omega_{\mathrm{atom}}$. The amount of squeezing is directly given by the Fourier transform of the two photon wavefunction,
\begin{eqnarray}
\lefteqn{\psi(\omega_1,\omega_2) = \sqrt{2} \Gamma^\prime \tan^2\left(\frac{\theta}{2}\right) \delta(\omega_1+\omega_2)}
\nonumber \\ &&
\times\left(\frac{1}{\Gamma^\prime +i(\omega_\beta-\omega_1)}+\frac{1}{2\Gamma^\prime +i(\omega_\beta+\omega_1)}\right).
\end{eqnarray}
Thus, the complete response of a single atomic system to an off-resonant coherent driving field can be described by the combination of a linear coherent shift and four wave mixing effects that result in sideband squeezing. More complicated responses are possible at higher input intensities, but the quantum state in the output can always be expressed in terms of a superposition of different quantum jump sequences with the appropriate linear shift of the coherent amplitude. The present analysis provides a detailed description of these quantum effects and can be used to test intuitive assumptions about the usefulness of optical nonlinearities for quantum technologies.

\section{Conclusions}
\label{sec:concl}

We have analyzed the interaction between coherent light and a single atom using a completely quantum mechanical description of light field propagation and field-atom interaction. By changing the representation of the field and the atomic system, we found that the interaction can be described by a combination of coherent shifts and a sequence of temporally correlated quantum jumps that result in the generation of additional photons in the effective vacuum represented by the initial coherent state. If the excitation of the atom is sufficiently low, as is typically the case for strongly detuned fields, the coherent shift represents the linear response of the atom and the quantum jumps correspond to photon scattering by four wave mixing. Significantly, the linear shift can be made arbitrarily strong by increasing the pulse duration. It may therefore be possible to realize a quantum controlled coherent shift of the light field, where the control qubit is encoded in the electronic states of the atomic system. Such a quantum controlled shift gate could then be used to generate cat-state superpositions of coherent states with negligible overlap. 

The main non-linear effect of the light-matter interaction can be expressed in terms of the generation of additional photon pairs, where one of the photons is resonant with the atom and the other photon is emitted at a frequency of $2 \omega_{\mathrm{light}}-\omega_{\mathrm{atom}}$. These photon pairs can also be explained as an effect of vacuum squeezing caused by four wave mixing between the input field and the vacuum fluctuations of the two sidebands. The dominant non-linear effect in the field-atom interaction can therefore be described in terms of a highly frequency dependent third order nonlinearity that results in particularly strong interactions between the input field and frequencies at the atomic resonance. Taking into account such resonant non-linearities may provide significant new insights into the possibilities and limitations of non-linear optical elements for quantum information processes. In particular, it is likely that the main practical difficulties of achieving effective single mode operations in non-linear quantum optical devices will originate from four wave mixing effects associated with the off-resonant atomic states involved in the non-linear response of the medium. 

Essentially, we have shown that the choice of an appropriate representation is crucial for the description of optical quantum effects involving a large number of photons. It may therefore be interesting to consider the application of our approach to more complicated systems, in order to identify the relevant physical effects in non-linear quantum optical devices. In particular, it seems to be possible to achieve a better intuitive understanding of the quantum dynamics involved in the off-resonant light-matter interaction, where a large number of photons is needed to achieve any significant effects. 

\section*{Acknowledgement}
This work was supported by JSPS KAKENHI Grant No. 24540427.

\end{document}